\documentclass[twocolumn,aps,prl,amssymb,showpacs]{revtex4}
\usepackage{graphicx}% Include figure files
\usepackage{dcolumn}% Align table columns on decimal point
\usepackage{bm}% bold math

\usepackage{amsmath}
\usepackage{dsfont}
\usepackage{amsthm}

\newcommand{\saddleint}{I}
\newcommand{\bathint}{J}
\newcommand{\R}{\mathds{R}}
\newcommand{\ud}{\mathrm{d}}
\newcommand{\ui}{\mathrm{i}}
\newcommand{\ue}{\mathrm{e}}
\newcommand{\pa}{\partial}
\newcommand{\la}{\langle}
\newcommand{\ra}{\rangle}
\providecommand{\abs}[1]{\lvert#1\rvert}
\newcommand{\cW}{{\mathcal W}}
\newcommand{\adM}{\operatorname{Mad}}

\newcommand{\rem}[1]{}

\newtheorem{platz}{{\bf Fig.}} % a dummy theorem to count figures!
\newcommand{\capsty}{\footnotesize}

\rem{
\newcommand{\Capts}[1]{#1}
\newcommand{\FIGo}[3]{%
\marginpar{ \begin{platz} \label{#1} ~ \end{platz} \vspace*{1.5ex} }}

} % end rem

\newcommand{\Capts}[1]{}
\newcommand{\FIGo}[3]{\begin{figure}[floatfix]%

#3% % comment this line out to remove all figures
\caption[]{\capsty #2}%
\label{#1}%
\end{figure}}
\begin{document}

\title{The efficient computation of transition state resonances and 
reaction rates from a quantum normal form}

\author{Roman Schubert}
%\email{Roman.Schubert@bris.ac.uk}
\author{Holger Waalkens}
%\email{H.Waalkens@bris.ac.uk}
\author{Stephen Wiggins}
%\email{S.Wiggins@bris.ac.uk}
\affiliation{School of Mathematics, University Walk, University of
Bristol, Bristol BS8 1TW, United Kingdom}

\date{\today}

\begin{abstract}
A quantum version of a recent formulation of transition state theory 
in {\em phase space} is presented. The theory developed provides an algorithm 
to compute quantum reaction rates and the associated Gamov-Siegert resonances with very high accuracy.  
The algorithm is especially efficient for multi-degree-of-freedom systems  
where other approaches are no longer feasible.

\end{abstract}

\pacs{82.20.Ln, 34.10.+x, 05.45.-a}

\maketitle

%%%%%%%%%%%%%%%%%%%%%%%%%%%%%%%%%%%%%%%%%%%%%%%%%%%%%%%%%%%%%%%%%%%%%%%%
{\em Introduction.---}
The question of how, as Marcus \cite{Marcus92} formulates it, a system ``skis the reaction slope'' 
is one of the crucial questions in reaction dynamics.
Experimental techniques like photodissociation of jet-cooled molecules, molecular beam experiments  or transition state spectroscopy 
give detailed information about the reaction process as has recently been demonstrated, e.g., for the `paradigm' 
reaction of Hydrogen atom-diatom collisions 
(see, e.g., the review paper \cite{SkodjeYang04}).
A chemical reaction can often be viewed as the scattering problem across a saddle point of the interaction potential. 
The cumulative reaction probability is then given by
\begin{equation}
N(E) = \mbox{tr } \widehat{t}\, \widehat{t}^T = \sum_{n_R\,n_P} |\langle
\psi_{\text{out}\,n_p} | \widehat{S}| \psi_{\text{in}\,n_r} \rangle|^2
\end{equation}
where $\widehat{t}$ is the transmission subblock of the scattering operator $\widehat{S}$ for energy $E$
and the summation in the latter expression runs over all incoming reactant
states with quantum numbers $n_r$  
and outgoing product states with quantum numbers $n_p$.
The {\em ab initio} quantum mechanical computation of $N(E)$ soon becomes very expensive if the number of atoms in the system increases beyond 3
and one has to resort to suitable approximations.
The main approach to compute $N(E)$ {\em classically} 
is transition state theory which was invented by Eyring, Polanyi and Wigner
in the 30's. 
The main idea 
is to define a dividing surface that divides the energy 
surface into a reactant and a product component and compute the rate
from the directional phase space flux through this surface.
In order not to overestimate the rate the
dividing surface must not be recrossed by reactive trajectories.
%to be a so called  `surface of no return.' This means
%that the dividing surface is crossed exactly once by reactive trajectories 
%and not crossed at all by non-reactive trajectories. 
In the 70's Pechukas, Pollak and others \cite{PechukasMcLafferty73} showed that for two degrees of freedom 
such a dividing surface can be constructed from a periodic orbit that leads to the so called 
\emph{periodic orbit dividing surface}.
Recently it has been shown that a generalization to higher dimension can be achieved from
a {\em normally hyperbolic invariant manifold} (NHIM) \cite{Wiggins90} ---  a fundamentally new object that 
takes the place of the periodic orbit.
\rem{
The problem of how to rigorously 
construct such a surface  was solved in the 70's for systems with two degrees of freedom
by Pechukas, Pollak and others \cite{PechukasMcLafferty73}. 
The basis for their construction is a periodic
orbit that leads to the so called `periodic orbit dividing surface.' 
Recently it has been shown how to generalize this to higher dimensions \cite{}. Here
a fundamentally new
object  --- a normally hyperbolic invariant manifold (NHIM)\cite{Wiggins90} ---  takes the
place of the periodic orbit. 
}%rem
The dynamics is controlled by the NHIM's stable and unstable manifolds 
which act as separatrices that divide the reactive trajectories from the nonreactive trajectories. 
%Recently, it has been shown how 
The NHIM, its stable and unstable manifolds and a dividing surface 
with the desired properties
can be directly constructed from an algorithm based on a Poincar{\'e}-Birkhoff normal form procedure 
\cite{UJPYW01}.

Much effort has been devoted to developing a quantum version of transition state theory whose 
implementation remains feasible 
for multi-dimensional systems (see the flux-flux autocorrelation function formalism by Miller and 
coworkers \cite{Miller98}).
In this Letter we present a quantum version of the normal form
procedure that lead to the construction of the high-dimensional phase space structures that 
govern the classical reaction dynamics and demonstrate that this {\em quantum normal form} approach to transition state theory
provides an efficient procedure to compute quantum reaction rates and the corresponding Gamov-Siegert resonances \cite{FriedmanTruhlar91}. 

%%%%%%%%%%%%%%%%%%%%%%%%%%%%%%%%%%%%%%%%%%%%%%%%%%%%%%%%%%%%%%%%%%%%%%%%%%%
{\em The quantum normal form.---}
The main idea of which the seed can already be found, e.g., in \cite{HernandezMiller93},
is to derive a local approximation of the Hamilton operator of the scattering problem that is valid near the saddle, and in order to facilitate further computations, takes a much simpler form than the original Hamiltonian.  
This can be achieved in a systematic way by a procedure based on the  Wigner-Weyl calculus that 
has been used by others before to compute energy spectra  associated with stable equilibria \cite{FriedEzra88}.
Here the manipulations of an operator $\widehat{A}$ are expressed in terms of its \emph{symbol} which 
is the function  $A(p,x)$ on the $f$-degrees-of-freedom  phase space $(p,x)\in\R^{2f}$ 
defined by
\begin{equation}
\widehat{A}\psi(x) = \frac{1}{(2\pi\hbar)^f}
\int_{\R^{2f}} \ue^{\frac{\ui}{\hbar}\la x-y,p\ra} A\bigg(p,\frac{x+y}{2}\bigg)\,
\psi(y)\,\ud y \ud p \,.
\end{equation}
Defining the multiplication $*$ of two symbols $A$ and $B$ according to
\begin{equation}\label{eq:star-product}
A*B  = A \exp\bigg(\ui\frac{\hbar}{2}[\la
  \stackrel{\leftharpoonup}{\pa}_{x},\stackrel{\rightharpoonup}{\pa}_{p}\ra-\la \stackrel{\rightharpoonup}{\pa}_{x},\stackrel{\leftharpoonup}{\pa}_{p}\ra]\bigg) B\,,
\end{equation}
where the arrows indicate whether the partial differentiation acts to the left (on $A$) or to the right (on $B$),
gives the property that the quantization of the product of two symbols $A$ and $B$  is equal to the product of the quantizations of the individual symbols, i.e. $\widehat{A}\widehat{B}=\widehat{A*B}$\,. The $*$-product leads to the definition of the Moyal bracket
\begin{equation}
\{A,B\}_M = \frac{\ui}{\hbar} (A*B-B*A)\,.
\end{equation}
We define the {\em order} $s$ of a monomial 
$p^{\alpha}x^{\beta}\hbar^n\equiv p_1^{\alpha_1}\cdots
p_f^{\alpha_f}x_1^{\beta_1}\cdots x_f^{\beta_f} \hbar^n$ according to 
$s=\abs{\alpha}+\abs{\beta}+2n\equiv \alpha_1+\dots+\alpha_f+\beta_1+\dots+\beta_f+2n$, and denote the vector space of 
polynomials spanned by monomials of order $s$ by $\cW^s$. 
For $A\in \cW^s$ and $B\in \cW^{s'}$ we define 
the Moyal adjoint by $\adM_AB:=\{A,B\}_M $, then  its iterates satisfy 
$[\adM_A]^nB\in \cW^{n(s-2)+s'}$.

We start from a Hamilton operator $\widehat{H}$
whose symbol  has expansion
\begin{equation}  \label{eq:Hstart}
H = E_0 + \sum_{s=2}^\infty H_s
\end{equation}
where $E_0$ is a constant energy  and $H_s\in \cW^s$. 
Like in the case of classical transition state theory 
\cite{UJPYW01}
we assume that
the second order term is of the form
\begin{equation} \label{eq:H2start}
H_2 = \lambda \saddleint + \omega_2 \bathint_2 + \dots + \omega_f \bathint_f
\end{equation}
with the integrals
\begin{equation} \label{eq:defints}
I= p_1 x_1 \,,\quad J_k = \frac12 \big( p_k^2 + x_k^2 \big)\,,\quad k=2,\dots,f\,.
\end{equation}
This corresponds to a classical equilibrium point of saddle-center-$\dots$-center stability type (`saddle' for short), 
i.e. the matrix associated with the linearized system has one pair of real eigenvalues 
$\pm \lambda$ associated with the saddle or `reaction coordinate'
and $f-1$ pairs of imaginary eigenvalues 
$\pm \ui \omega_k$, $k=2,\dots,f$,  associated with the center or `bath' degrees of freedom. 
Note that $p_1x_1=(\tilde{p}_1^2-\tilde{q}_1^2)/2$ where $(p_1,x_1)$ and  $(\tilde{p}_1,\tilde{q}_1)$ 
are related by a rotation of 45$^\circ$.
We restrict ourselves to the generic non-resonant case where the frequencies $\omega_k$, $k=2,\dots,f$, are rationally independent

In order to simplify the Hamiltonian $\widehat{H}$ we will transform it by  successive
conjugations with 
unitary operators, $\widehat{H}=:\widehat{H}^{(2)}\to\widehat{H}^{(3)}\to\widehat{H}^{(4)}\to \cdots \to\widehat{H}^{(N)}$, 
where 
\begin{equation} \label{eq:UHUt}
\widehat{H}^{(n)}= 
\ue^{\ui \widehat{W}_n/\hbar } \widehat{H}^{(n-1)} \ue^{ - \ui \widehat{W}_n /\hbar}  \,\, ,
\end{equation}
with $W_n\in \cW^n$. Using the Moyal adjoint  the symbol of  the right hand side  can be expanded as
\begin{equation}\label{eq:expanded}
H^{(n)}=\sum_{k=0}^{\infty} \frac{1}{k!} [\adM_{W_n}]^k H^{(n-1)}\,\, .
\end{equation}
If we expand furthermore each of the symbols $H^{(n)}$ in a generalized Taylor series as in \eqref{eq:Hstart},
$H^{(n)}=E_0+\sum_{s\geq 2}H_s^{(n)}$, with $H_s^{(n)}\in W^s$, then using \eqref{eq:expanded}
the terms in these series can be  related by 
\begin{equation}
H_s^{(n)}=\sum_{k=0}^{[\frac{s-2}{k-2}]} \frac{1}{k!}[\adM_{W_n}]^k H^{(n-1)}_{s-n(k-2)}\,\, ,
\end{equation}
where $[\cdot]$ denotes the integer part. Notice that for $s<n$,  
$H_s^{(n)}=H_s^{(n-1)}$, and for $s=n$ we obtain 
\begin{equation}\label{eq:hom}
H_n^{(n)}=H_n^{(n-1)}+\{W_n,H_2\}
\end{equation}
where we have used that $H^{(n)}_2=H_2$ for all $n\geq 3$ and that the Moyal bracket 
reduces to the Poisson bracket $\{\cdot, \cdot\}$ if one of the functions is quadratic. This is the 
homological equation which is familiar from the classical normal form algorithm, \cite{UJPYW01}, 
and under the non-resonance conditions on the frequencies $\omega_k$, given any $H_n^{(n-1)}\in \cW^n$ 
there exists a unique $W_n \in \cW^n$ such that $H_n^{(n)}$ can be written  as a function 
of the actions $I, J_2,\cdots ,J_f$ alone (or, equivalently, $H_n^{(n)}$ satisfies  
$\{I,H_n^{(n)}\}=\{J_2, H_n^{(n)}\}=\cdots = \{J_f, H_n^{(n)}\}=0$).

Choosing the generators of the unitary transformations $W_n$ recursively for 
$n=3,4,\ldots N$ as solutions of \eqref{eq:hom} we obtain an operator 
$\widehat{H}^{(N)}$ whose symbol is of the form 
$H^{(N)}=E_0+\sum_{s=2}^{N}H_s^{(N)} +R^{(N+1)}$ where the first part is a polynomial  
in  the actions $I, J_2,\cdots J_f$, i.e., is in \emph{normal form}, and the remainder 
$R^{(N+1)}$ consists of terms of order $N+1$ and higher. In a final step 
we want to express the quantization of the normal form part as an operator function 
of the quantized actions $\hat{I}$ and $\widehat{J}_k$. To this end we use a recursion 
relation for $\widehat{I^n}$ (and a similar one for the $J_k$), 
$\widehat{I^{n+1}}=\widehat{I}\widehat{I^n} -  \widehat{I^{n-1}} n^2 \hbar^2/4$, which can be 
derived from the product formula \eqref{eq:star-product}. This allows us to express 
quantizations of powers of $I$ as a polynomial in powers of $\widehat{I}$. In this way we
find a polynomial $H^{(N)}_{\text{QNF}}$ such that 
\begin{equation}
U_N^*\widehat{H}U_N=H^{(N)}_{\text{QNF}}(\widehat{I}, \widehat{J}_2,\cdots ,\widehat{J}_f)
+\widehat{R}^{(N+1)}\,\, ,
\end{equation}
where $U_N=\prod_{n=3}^N \ue^{-\ui \widehat{W}_n/\hbar}$.
$H^{(N)}_{\text{QNF}}(\widehat{I}, \widehat{J}_2,\cdots ,\widehat{J}_f)$ 
is called the \emph{quantum normal form} (QNF) of $\widehat{H}$ of order $N$. 
The remainder term $\widehat{R}^{(N+1)}$ has a symbol which is of order $N+1$ and is therefore 
very small near the saddle point. Hence  the dynamics near the saddle point can be described with high accuracy 
by the QNF Hamiltonian. The advantage of the QNF Hamiltonian is that it is an operator function 
of the commuting operators $\widehat{I}$ and $\widehat{J}_k$ whose properties are well understood. 
Finally we point out that in the limit $\hbar\to 0$ we recover the classical normal form of order $N$, 
$\lim_{\hbar\to 0}H^{(N)}_{\text{QNF}}(I,J_2,\cdots, J_f)=H^{(N)}_{\text{cl}}$.  

We implemented the algorithm of the  QNF computation 
in the programming language {\sf C++}.

\def\figeckart{%
Transmission probability $T(E)$  and Gamov-Siegert resonances in the complex energy plane for the Eckart potential.
The parameters are $a=1$, $10 \, A=B=5$ for the Eckart potential, and  $m=1$ and $\hbar=0.1$.
}
\def\FIGeckart{
\centerline{
\includegraphics[angle=0,width=6.6cm]{fig1}%Figures/eckart_as_cumulative_xfig
}
}
\FIGo{fig:eckart}{\figeckart}{\FIGeckart}
%%
%%%%%%%%%%%%%%%%%%%%%%%%%%%%%%%%%%%%%%%%%%%%%%%%%%%%%%%%%%%%%%%%%%%%%%%
{\em Resonances and reaction rates.---}
The eigenfunctions of the truncated QNF Hamiltonian are tensor products of harmonic
oscillator wave functions for the center degrees of freedom $x_k$, 
$k=2,\dots,f$, and eigenfunctions of the operator
\begin{equation}
\widehat{\saddleint} = \frac{\hbar}{\ui}\bigg(x_1\pa_{x_1}+\frac{1}{2}\bigg)
\end{equation}
associated with the saddle direction. The operator $\widehat{\saddleint}$ has
eigenfunctions  \cite{ColindeVerdiereParisse94}
\begin{equation}
\psi_{\text{out}\,\pm}(x_1 )= \Theta(\pm x_1)\abs{x_1}^{-1/2+\ui \saddleint/\hbar}\,\, ,
\end{equation}
$\Theta$ being the step function, which are outgoing waves.
Incoming waves can be defined from the Fourier transforms
\begin{equation} \label{eq:psiin}
\begin{split}
\psi_{\text{in}\, \pm}(x_1)=\frac{1}{\sqrt{2\pi\hbar}} \int {\psi}_{\text{out}\, \mp}^*(y_1)
\ue^{\frac{\ui}{\hbar} x_1 y_1}\,\, \ud y_1 \, .
\end{split}
\end{equation}
Note that in order to classify the eigenfunctions as `outgoing' or `incoming' it is convenient 
to rotate the coordinates $(p_1,x_1)$ back to the more standard notation $(\tilde{p}_1,\tilde{q}_1)$
for a potential barrier mentioned above.
Expressing the functions $\psi_{\text{in}\, \pm}$ in terms of 
$\psi_{\text{out}\, \pm}$ gives the entries of a `local S-matrix',
\begin{eqnarray} \label{eq:psiinplus}
\psi_{\text{in}\, +} &=& S_{n\, 11}\, \psi_{\text{out}\,+} +  S_{n\, 12}\, \psi_{\text{out}\,-}\,,\\
\label{eq:psiinminus}
\psi_{\text{in}\, -} &=& S_{n\, 21}\, \psi_{\text{out}\,+} +  S_{n\, 22}\, \psi_{\text{out}\,-}\,.
\end{eqnarray}
Here $n$ denotes the vector of nonnegative integers $(n_2,\dots,n_f)$ formed by 
the quantum numbers of the modes in the center directions. 
The local S-matrix is block diagonal due to the separability of the QNF.  Mode mixing is a `global' effect which occurs from connecting the local wave functions to the
asymptotic reactants and products wave functions.
However, the local S-matrix alone already contains the full information needed to 
compute reaction rates and resonances. 

Evaluating 
the integrals (\ref{eq:psiin}) gives
\begin{equation}
S_{n}(E)=\frac{\ue^{\ui (\frac{\pi}{4}-\frac{\saddleint}{\hbar}\ln\hbar )}}{\sqrt{2\pi}}
\Gamma\bigg(\frac{1}{2}-\ui\frac{\saddleint}{\hbar}\bigg)
\begin{pmatrix} 
-\ui \ue^{-\frac{\pi}{2}\frac{\saddleint}{\hbar}} & \ue^{\frac{\pi}{2}\frac{\saddleint}{\hbar}}  \\
\ue^{\frac{\pi}{2}\frac{\saddleint}{\hbar}} & -\ui \ue^{-\frac{\pi}{2}\frac{\saddleint}{\hbar}} 
\end{pmatrix}
\end{equation}
with $\saddleint$ being implicitly defined by 
\begin{equation}\label{eq:Iimplicit}
H^{(N)}_{\text{QNF}}
\big( \saddleint,\hbar(n_2+1/2), \dots ,\hbar(n_{f}+1/2) \big)= E\,.
\end{equation}
The transmission probability of mode $n$ is 
\begin{equation}
T_n(E)=\abs{S_{n\,12}(E)}^2 
=(1+\ue^{-2\pi \frac{\saddleint}{\hbar}})^{-1}\,,
\end{equation}
which gives the cumulative reaction probability 
$N(E) = \sum_n T_n(E)$.
The S-matrix has poles at $\saddleint = -\ui \hbar(n_1 + 1/2)$ for nonnegative integers $n_1$ 
and these define the Gamov-Siegert resonances via \eqref{eq:Iimplicit}.

%%%%%%%%%%%%%%%%%%%%%%%%%%%%%%%%%%%%%%%%%%%%%%%%%%%%%%%%%%%%%%%%%%%%%%%%%
{\em Examples.---}
We at first illustrate the procedure for 1D potential barriers, i.e.  Hamiltonians of the form
$
H = p^2/(2m) + V(x)\,,
$
where $V(x)$ has a maximum which we can assume to be at $x=0$.
The second order QNF is easily obtained and gives the well know result 
$\widehat{H}_{\text{QNF},\,2}=V(0) + \lambda \widehat{I}$ with $\lambda=(-V''(0)/m)^{1/2}$ which
is equivalent to approximating the potential barrier by an inverted parabola. 
The first nontrivial correction to this result 
comes from the fourth order QNF $\widehat{H}_{\text{QNF},\,4}$ given by
\begin{equation}
\begin{split}
& V(0) + \lambda \widehat{I}  + \frac{1}{16 m^2 \lambda^2}
 \left( \frac{5}{3 m \lambda^2}V'''(0) +  V^{IV}(0)  \right) \widehat{I}^2 \\
& -  \frac{1}{64 m^2\lambda^2}
\left( \frac{7}{ 9 m \lambda^2}V'''(0) +
 V^{IV}(0)  \right)\hbar^2\,.
\end{split}
\end{equation}
\def\figeckarterrors{%
(a) Errors for the transmission probability of the  Eckart potential computed from the QNF. 
(b) Errors for the QNF resonances 
(differences between the corresponding QNF and exact complex energies) 
of the Eckart potential as a function of the  quantum number $n_1$. 
The different colors correspond to different orders of the QNF.
The parameters for the Eckart potential are the same as in Fig.~\ref{fig:eckart}.
}
\def\FIGeckarterrors{
\centerline{
\includegraphics[angle=0,width=4.2cm]{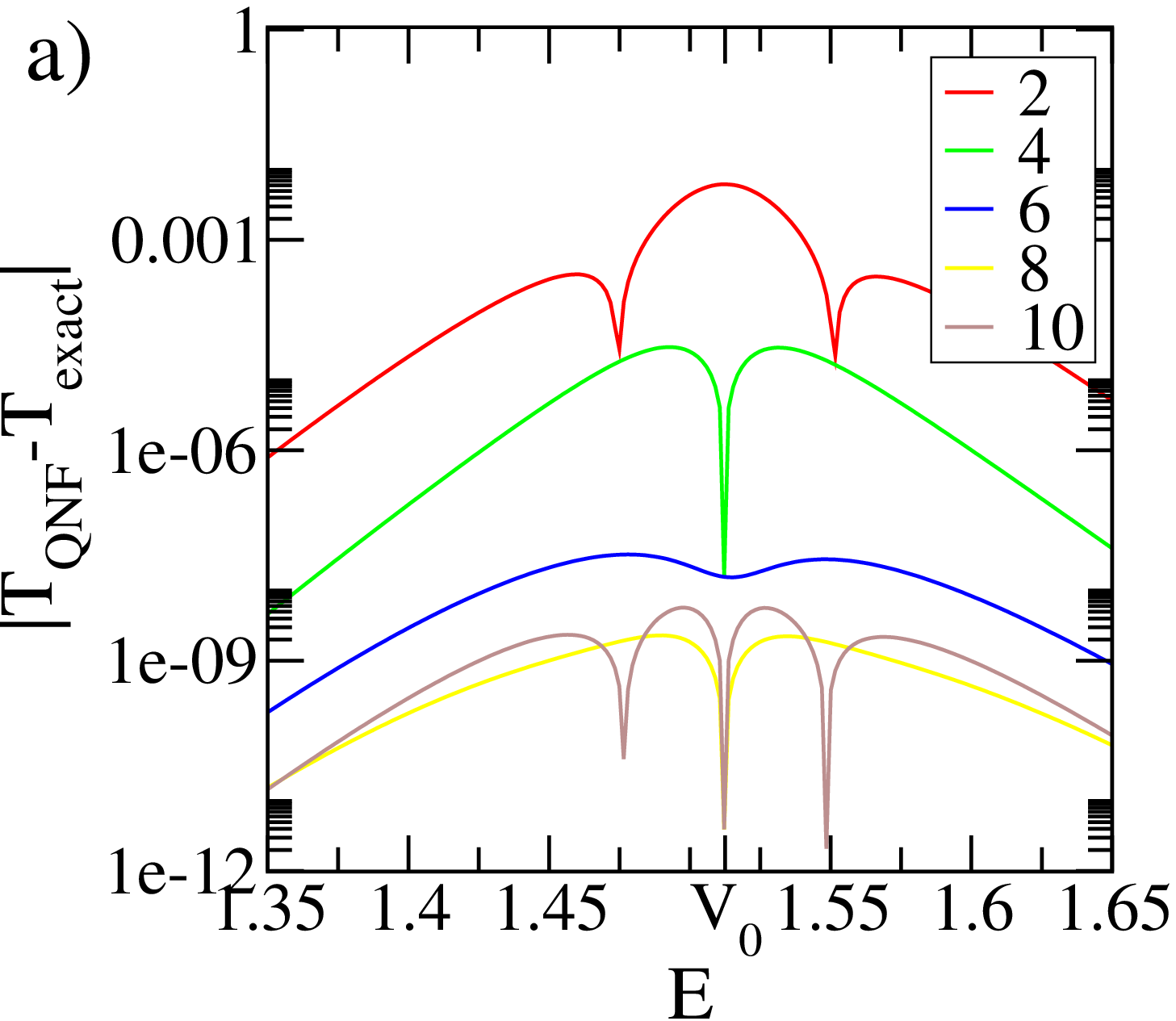}%{Figures/cummulative_eckart_as_difference_big_xfig}
\includegraphics[angle=0,width=4.2cm]{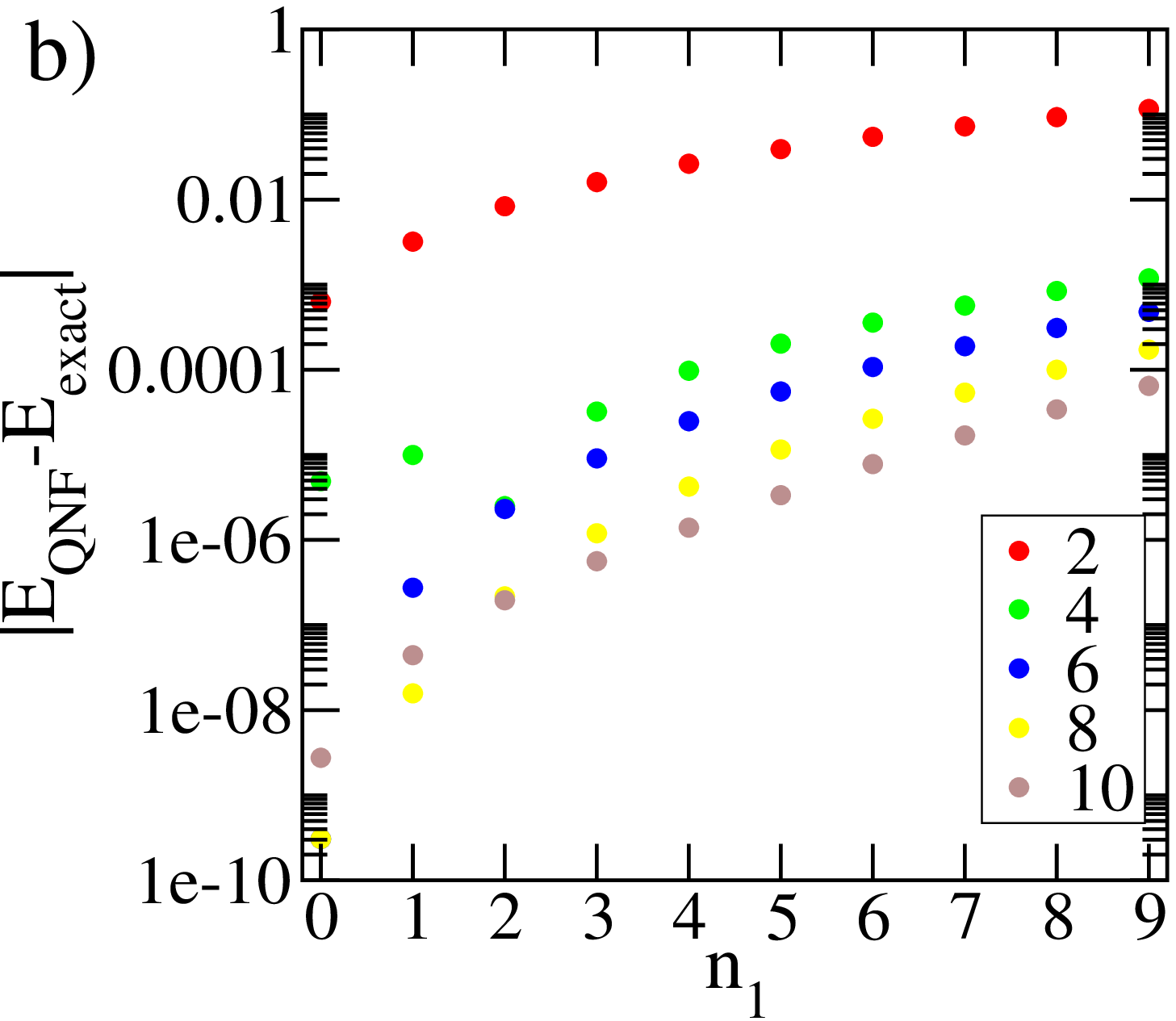}%{Figures/eckart_as_poles_difference_big_xfig}
}
}
\FIGo{fig:eckarterrors}{\figeckarterrors}{\FIGeckarterrors}
We apply the QNF to the Eckart potential \cite{Eckart30} 
\begin{equation}
V_{\text{Eckart}}(x) = A \frac{\ue^{(x+x_0)/a}}{1+\ue^{(x+x_0)/a}} + B \frac{\ue^{(x+x_0)/a}}{(1+\ue^{(x+x_0)/a})^2}
\end{equation}
with $x_0=a\ln (B+A)/(B-A)$ and $B>A\ge0$.
Figure~\ref{fig:eckart} shows the exact transmission probability which is known analytically, and the
exact string of resonances together with the resonances from the second order QNF which
have constant imaginary part. 
The bending of the string of exact resonances 
is a nonlinear effect that is very well described already by the fourth order QNF. 
The excellent accuracy of the resonances and the cumulative reaction probability computed from higher orders of the QNF
is illustrated in Fig.~\ref{fig:eckarterrors}. 

\def\figeckartmorsecoupled{%
(a) The top panel shows the cumulative reaction probability $N(E)$ (oscillatory curve) and the classical flux divided by $2\pi \hbar$ (smooth curve)  
for  the Eckart-Morse potential defined in the text with $\epsilon =0$. The bottom panel shows the resonances in the complex energy plane 
marked by circles for the uncoupled case $\epsilon =0$ and by crosses for the strongly coupled case  $\epsilon = 0.3$.
For the coupled case the numerically exact resonances are computed from the complex dilation method 
\cite{Moiseyev98}. 
 The parameters for the Eckart potential are the same as in
Fig.~\ref{fig:eckart}. The parameters for the Morse potential are $D_e=1$
$\alpha=1$. Again we choose $m=1$ and $\hbar=0.1$.
(b) Errors for the cumulative reaction probability in (a) for different orders of the QNF.
(c) Average errors  $<|E_{\text{QNF}}-E_{\text{exact}}|>_{n_1} =\sum_{n_2=0}^5 |E_{\text{QNF}}(n_1,n_2) - E_{\text{exact}}(n_1,n_2)|/6$
for the QNF resonances in (a) for the coupled case $\epsilon = 0.3$ 
as a function of the quantum numbers $n_1$.
}
\def\FIGeckartmorsecoupled{
\centerline{
\includegraphics[angle=0,width=8.3cm]{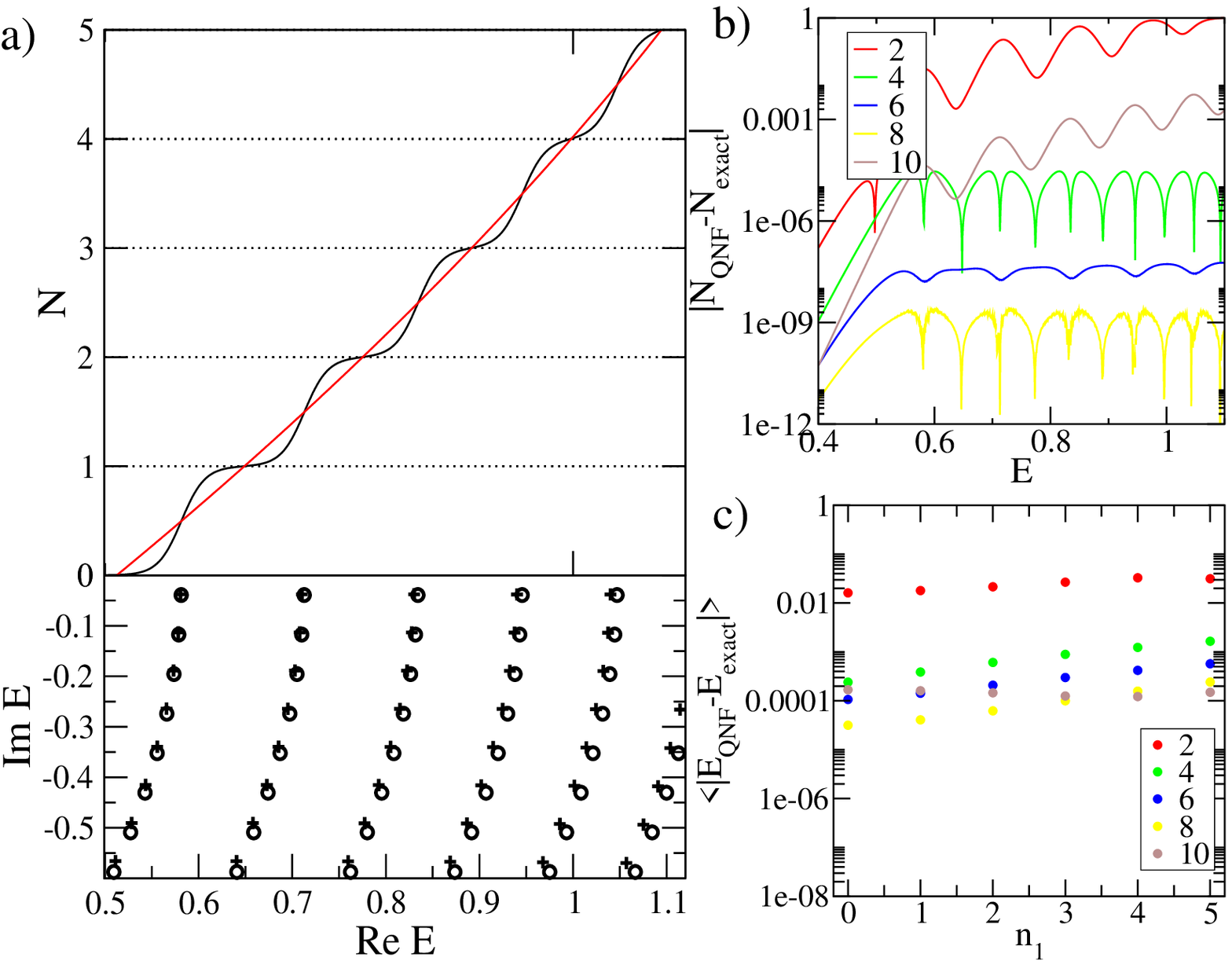}%{Figures/cum_res_eckart_as_morse_xfig}
}
}
\FIGo{fig:eckartmorsecoupled}{\figeckartmorsecoupled}{\FIGeckartmorsecoupled}

We next consider the two-degree-of-freedom example of a Hamiltonian 
with an Eckart potential in the $x_1$-direction, a Morse potential 
\begin{equation}
V_{\text{Morse}}(x_2) =  D_e\left( \ue^{-2\alpha x_2} - 2 \ue^{-\alpha x_2} \right)
\end{equation}
in the $x_2$-direction,
plus a kinetic coupling $\epsilon p_1 p_2$. In Fig.~\ref{fig:eckartmorsecoupled} the cumulative reaction probability and the resonances computed from the QNF are 
compared with the exact results. In the uncoupled case $\epsilon =0$, $N(E)$  increases as a function of $E$ 
at integer steps each time a new transition channel opens, i.e. when the transmission probability $T_{n_2}(E)$  of a Morse oscillator mode $n_2$ switches from 0 to 1. 
For both the uncoupled and  strongly coupled case the resonances form a distorted lattice parametrized by the mode quantum number 
$n_2$ in horizontal direction and the quantum number $n_1$ in vertical direction. 
Similar to the 1D example, each string of constant $n_2$ is related to one step of $N(E)$.

Like in the 1D case 
the agreement of the QNF results with the exact results is excellent and this remains the case even for the strongly coupled system.
The QNF is an asymptotic expansion which in general does not converge. 
For the system shown in the example this can be seen from the fact that the 10th order QNF does not lead to an improvement of the results obtained from the 
8th order QNF.

%%%%%%%%%%%%%%%%%%%%%%%%%%%%%%%%%%%%%%%%%%%%%%%%%%%%%%%%%%%%%%%%%%%%%%%%
{\em Conclusions.---}
The QNF computation of reaction probabilities and the corresponding Gamov-Siegert resonances is highly promising since it opens the way to study high dimensional systems for which other techniques based on the {\em ab initio} 
solution of the quantum scattering 
problem like the complex dilation method \cite{Moiseyev98}  or the utilization of an absorbing potential \cite{NeumaierMandelshtam01} are no longer feasible. 
In fact, the numerical effort for computing the QNF is only slightly higher than the effort for computing  the classical normal form (NF).
The main difference is that the Poisson bracket in the classical NF needs to be replaced by the Moyal bracket. 
The storage of the QNF polynomials also requires only
a disk space similar to the classical NF. 
Moreover, the QNF gives an explicit formula for the resonances from which they 
can be computed directly by inserting the corresponding quantum numbers.
This leads to a direct assignment of the resonances. 
The QNF provides a quantum version of transition state theory that, in the semiclassical limit, is in accord with the 
classical phase structures that govern the reaction dynamics. 
In fact, the classical phase space structures form the skeleton for the scattering and resonance wavefunctions, and 
exploiting this relationship which will give a deep insight into the now experimentally accessible quantum reaction dynamics
is the subject of our future studies (see also \cite{Creagh04}).

This work was supported by the Office of Naval Research, EPSRC and the Royal
Society.

%%%%%%%%%%%%%%%%%%%%%%%%%%%%%%%%%%%%%%%%%%%%%%%%%%%%%%%%%%%%%%%%%%%%%%%%

%\bibliography{/home/mazhw/Tex/Bibs/extern,/home/mazhw/Tex/Bibs/fg4}

\end{document}